\documentclass[11pt]{article}
\usepackage{geometry}                
\usepackage{graphicx}
\usepackage{amssymb}
\usepackage{amsmath}
\usepackage{amsfonts}
\usepackage{graphicx}
\usepackage{epstopdf}
\usepackage{color}
\usepackage{cite}

\newcommand{\be}{\begin{equation}}
\newcommand{\ee}{\end{equation}}
\newcommand{\bea}{\begin{eqnarray}}
\newcommand{\eea}{\end{eqnarray}}
\newcommand{\beas}{\begin{eqnarray*}}
\newcommand{\eeas}{\end{eqnarray*}}

\newcommand{\dd}{\mathrm{d}}

\usepackage{float}
\usepackage[titletoc,title]{appendix}

\usepackage{parskip}

\title{Optimizing Expected Shortfall under an $\ell_1$ constraint - an analytic approach}
\date{}
\author{G\'abor Papp$^1$, Imre Kondor$^{2,3,4}$ and Fabio Caccioli$^{5,6,3}$\\
{\it 1- E\"otv\"os Lor\'and University, Institute for Physics, Budapest, Hungary } \\
{\it 2-Parmenides Foundation, Pullach, Germany} \\
{\it 3- London Mathematical Laboratory, London, UK} \\
{\it 4- Complexity Science Hub, Vienna, Austria} \\
{\it 5- University College London, Department of Computer Science,} \\
{\it London, WC1E 6BT, UK} \\
{\it 6- Systemic Risk Centre, London School of Economics and Political Sciences, London, UK}\\
}

\begin{document}
\bibliographystyle{unsrt}

\maketitle

\begin{abstract} 
Expected Shortfall (ES), the average loss above a high quantile, is the current financial regulatory market risk measure. Its estimation and optimization are highly unstable against sample fluctuations and become impossible above a critical ratio $r=N/T$, where $N$ is the number of different assets in the portfolio, and $T$ is the length of the available time series. The critical ratio depends on the confidence level $\alpha$, which means we have a line of critical points on the $\alpha-r$ plane. The large fluctuations in the estimation of ES can be attenuated by the application of regularizers. In this paper, we calculate ES analytically under an $\ell_1$ regularizer by the method of replicas borrowed from the statistical physics of random systems. The ban on short selling, i.e. a constraint rendering all the portfolio weights non-negative, is a special case of an asymmetric $\ell_1$ regularizer. Results are presented for the out-of-sample and the in-sample estimator of the regularized ES, the estimation error, the distribution of the optimal portfolio weights and the density of the assets eliminated from the portfolio by the regularizer. It is shown that the no-short constraint acts as a high volatility cutoff, in the sense that it sets the weights of the high volatility elements to zero with higher probability than those of the low volatility items. This cutoff renormalizes the aspect ratio $r=N/T$, thereby extending the range of the feasibility of optimization. We find that there is a nontrivial mapping between the regularized and unregularized problems, corresponding to a renormalization of the order parameters. \end{abstract}

\section{Introduction}

A risk measure is a functional on the probability distribution of the fluctuating returns of a security or a portfolio. Since it is impossible to condense all the information in a probability distribution into a single number, there is no unique way to choose the ``best'' risk measure. In Markowitz's ground breaking portfolio selection theory \cite{Markowitz1952Portfolio}, with the assumption of Gaussian distributed returns, variance offered itself as the natural risk measure. The crises of the late eighties and early nineties led both the industry and regulators to realize that the most dangerous risk lurked in the asymptotically far tail of the return distribution. To grasp this risk, a high quantile of the profit and loss distribution called Value at Risk (VaR) was introduced by J.P. Morgan \cite{Morgan1995Riskmetrics}. For a certain period VaR became a kind of industry standard, and it was embraced by international financial regulation as the official risk measure in 1996 \cite{Basel1996Overview}. Value at Risk is a threshold which losses only exceed with a small probability (such as e.g. 0.05 or 0.01), corresponding to a confidence level of                  $\alpha= 0.95$, resp. $0.99$. (In this context it is customary to regard losses as positive and profits as negative). As a quantile, VaR is not sensitive to the distribution of losses above the confidence level and is not subadditive when two portfolios are combined. This triggered a search for alternatives and led Artzner et al. \cite{Artzner1999Coherent} to formulate a set of axioms that any coherent risk measure should satisfy. The simplest and most intuitive of these coherent measures is the Expected Shortfall (ES) \cite{Acerbi2002Expected,Pflug2000Some}. ES is essentially the expected loss above a high quantile that can be chosen to be the VaR itself. After a long debate about the relative merits and drawbacks of ES, whose details are not pertinent to our present study, regulators adopted ES as the current official market risk measure to be used to assess the financial health of banks and determine the capital charge they are required to hold against their risks. The regulators and the industry settled on a confidence level of $\alpha=0.975$ \cite{Basel2016Minimum}.

ES is mainly designed to be a diagnostic tool. At the same time it is also a constraint that banks have to respect when considering the composition of their portfolios. It is then in their best interest to optimize ES, in order to keep their capital charge as low as possible. However, the optimization of ES is fraught with problems of estimation error, which is quite natural if one considers that the number of different items $N$ in a bank's portfolio can be very large, whereas the number of observations (the length of the available time series $T$) is always limited. In addition, at the regulatory confidence level one has to throw away 97.5\%
of the data. Moreover, the estimation error increases with the ratio $r=N/T$ and at a critical value of $r$ it actually diverges, growing beyond any limit. As shown in \cite{Kondor2010Instability} the instability of the optimization of ES (as well as all the coherent risk measures) follows directly from the coherence axioms \cite{Artzner1999Coherent}.

The divergence of ES is the signature of a phase transition. The critical $r$ for variance is $r=1$, for ES it is smaller or equal to $1/2$, its value depending on the confidence level $\alpha$. For ES there is then a line of critical points, a phase diagram, on the  $r-\alpha$ plane. A part of this phase diagram has been traced out by numerical simulations in \cite{Kondor2007Noise}, while the full phase diagram has been determined by analytical calculations by Ciliberti et al. \cite{Ciliberti2007On}. Going beyond merely determining the phase diagram, a detailed study of the estimation error and other relevant quantities has been performed inside the whole feasibility region in \cite{Kondor2015Contour, Caccioli2018Portfolio}, and it was shown that, due to the nontrivial behaviour of the contour lines of constant estimation error especially in the vicinity of $\alpha=1$, the number of data necessary to have a reasonably low estimation error was way above any $T$ available in practice.

Because of the large sample fluctuations of ES, its optimization constitutes a problem in high dimensional statistics \cite{Buhlmann2011Statistics}. A standard tool to tame these large fluctuations is to introduce regularizers, which penalize large excursions. Although the introduction of these penalties may seem an arbitrary statistical trick coming from outside of finance, it was shown in \cite{Caccioli2013Optimal} that these regularizers express liquidity considerations, and take into account, already at the construction of the portfolio, the expected market impact of a future liquidation. The regularizers are usually chosen to be some constraints on the norm of the portfolio weights. In \cite{Papp2016Variance} we studied the effect of an $\ell_2$ regularizer on ES and found that $\ell_2$ obviously suppresses the instability and, for sufficiently small $r$ and with a strong enough regularizer, it extends the range where the estimation error is reasonably small by a factor of about 4. 

It is interesting to see how an $\ell_1$ regularizer works with ES. (The importance of studying the effect of various regularizers in combination with the different risk measures was emphasized by \cite{Still2010Regularizing}). The regularizer $\ell_1$ is known to produce sparse solutions, which means that in order to rein in large fluctuations it eliminates some of the securities from the portfolio. This obviously contradicts the principle of diversification, but considerations of transaction costs or the technical difficulties of managing large portfolios may make it desirable to weed out the most volatile items from the portfolio, and this is precisely what a no-short constraint tends to do. 

It has been known for 20 years now that the optimization of ES can be translated into a linear programming problem \cite {Rockafellar2000Optimization}. Accordingly, as it has been realized in \cite{Caccioli2016Lp}, the piece-wise linear $\ell_1$ with an infinite slope corresponding to an infinite penalty on short selling can completely remove the instability of ES.
The purpose of this paper is to determine the effect of  $\ell_1$-regularization on the phase diagram and also on the behaviour of the various quantities of interest inside the region where the optimization of ES is feasible and meaningful. (We will see that as a result of regularization new characteristic lines appear on the $r-\alpha$ plane, beyond which the optimization of ES is still mathematically feasible, but the results become meaningless, as they correspond to negative risk.) In \cite{Caccioli2018Portfolio} a detailed analytical investigation of the behaviour of the estimation error, the in-sample cost, the sensitivity to small changes in the composition of the portfolio and the distribution of optimal weights were carried out in the non-regularized case. Here, we derive the same quantities for an $\ell_1$-regularized ES, including the special case where short selling is banned, that is when the portfolio weights are constrained to be non-negative. The density of the items eliminated from the portfolio, to be referred to as the ``condensate'' in the following, is also determined. The most striking result of the present study is that the regularized solution can be mapped back onto the unregularized one. We are not aware of a similarly tight relationship between a regularized and an unregularized problem, not only in a finance context, but neither in the general context of machine learning.

\section{Method and preliminaries}

If the true probability distribution of returns were known, it would be easy to calculate the true value of Expected Shortfall and the optimal portfolio weights. However, the true distribution of returns is unknown. In practice, one has to resort to either parametric, or historical estimations. In this paper, we will focus on historical estimation. This means one observes $N$ time series of length $T$ and estimates the optimal weights and ES on the basis of this information. It is clear that the weights and ES so obtained will deviate from their ``true'' values. (The latter would be obtained in an infinitely long stationary sample.) The deviation of the estimated values will be the stronger the shorter the length $T$ and the larger the dimension $N$. Performing this measurement on different samples one would obtain different estimates: there is a distribution of ES and of the optimal weights over the samples.

In a real market, one cannot repeat such an experiment multiple times. Instead, one has to squeeze out as much information as possible from a single sample of limited size. There are well-known numerical methods for this, like cross-validation or bootstrap \cite{Hastie2008Elements}. In contrast, in the present work we aim to obtain \emph{analytic} results. In order to mimic historical estimation, we choose a simple data generating process, such as a multivariate Gaussian. The true value of ES is easy to obtain for this case, which provides a standard to measure finite sample deviations from. Then we determine ES for a large number of random samples of length $T$ drawn from this underlying distribution,  average it over the random samples and finally compare this average to its true value. This procedure will give us an idea about how large the estimation error is for a given dimension $N$, sample size $T$ and confidence level $\alpha$, under the idealized conditions of stationarity and Gaussian fluctuations, and how much it will be reduced when we apply an $\ell_1$ regularizer of a given strength. It is reasonable to assume that the estimation error obtained under these idealized circumstances will be a lower bound to the estimation error for real-life processes. 

Now we wish to implement this program via analytic calculations. The averaging over the random samples just described is analogous to the averaging over the random realization of disorder in the statistical physics of random systems, which enables us to borrow methods from that field, in particular the replica method \cite{mezard1987Spin}. It assumes that both $N$ and $T$ are large, with their ratio $r=N/T$ kept finite (thermodynamic or Kolmogorov limit). A small value of $r$ corresponds to the classical setup in statistics where one has a large number of observations relative to the dimension. Estimates in this case are sharp and close to their true values. In contrast, when $r$ is of order unity, or larger, we are in the high dimensional limit where fluctuations are large. It is here that the regularizer becomes important.

In the usual application of $\ell_1$ in finite dimensional numerical studies, it eliminates the dimensions one by one, in a stepwise manner, as the strength of the regularizer is increasing. In our present work, the large $N,T$ limit and the averaging over infinitely many samples result in a continuous dependence of the ``condensate'' density, that is the relative number $N_0/N$ of the dimensions eliminated by $\ell_1$, on the aspect ratio $r$, the confidence level $\alpha$ and the strength of $\ell_1$. In a study of $\ell_1$-regularized variance \cite{Kondor2019Variancewithl1} we found that the stepwise increase of the density of eliminated weights in a numerical experiment nicely follows the continuous curve obtained analytically. It is obvious that the situation is similar in the case of ES as we have confirmed it by numerical simulations.
 
For the sake of simplicity, we will also assume that the returns are independent, that is the true covariance matrix is diagonal. This is not an innocent assumption: it will be seen, for example, that the maximum degree of sparsity that $\ell_1$ can achieve in this scheme is one half of the total number of dimensions, whereas for correlated returns the maximum sparsity can be either larger or smaller than $1/2$, according to whether correlations are predominantly positive or negative. Combining $\ell_1$ with a non-diagonal covariance matrix poses additional technical difficulties that we wish to avoid in the present account. However, we do allow the diagonal elements $\sigma_i$ of the covariance matrix to be different from each other.

As a further simplification, we do not impose any other constraint on the optimization of ES beside the budget constraint and the $\ell_1$ regularizer. In particular, we do not set a constraint on the expected return, and seek the global minimum of the regularized ES. This is in line with a number of studies, \cite{Kempf2006Estimating, Basak2009Jackknife, Frahm2010Dominating} among others, which focus on the global minimum in the problem of variance optimization, because of the extremely noisy estimates of the expected return. Furthermore, the global minimum is precisely what one needs in minimizing tracking-error, that is when trying to follow, say, a market index as closely as possible \cite{Basak2009Jackknife}.

The replica method used below have already been applied with minor variations to various portfolio optimization problems in a number of papers \cite{Ciliberti2007On, Ciliberti2007Risk, Vargahaszonits2008Instabilityofdownside, Caccioli2013Optimal, Caccioli2016Lp, Kondor2015Contour, Caccioli2016Lp, Shinzato2017Minimal, Kondor2017Variancenoshort, Caccioli2018Portfolio, Kondor2019Variancewithl1}, where the replica derivation of the main formulae were repeatedly explained, so we do not need to go through that exercise again here. Then the natural starting point for our present work is the detailed study of the behaviour of ES \emph{without} regularization in \cite{Caccioli2018Portfolio}. The argument there leads to a relationship between ES and an effective cost or free energy per asset $f$ as follows:

\be
\label{equationESCost}
  {\rm ES} = \frac{f r}{1-\alpha} \,.
\ee

The free energy $f$ itself is given by the minimum of a functional depending on six order parameters

\bea
\label{free_energy}
f({\lambda},{\epsilon},{q}_0,\Delta, {\hat{q}}_0,\hat{\Delta})&=&
{\lambda} +\frac{1}{r} (1-\alpha)\epsilon -\Delta{\hat{q}}_0-\hat{\Delta}{q}_0\\
\nonumber &+& \langle {\rm min}_w \left[V(w,z,\sigma)\right]\rangle_{\sigma,z}
 +\frac{\Delta}{2r\sqrt{\pi}}\int\limits_{-\infty}^{\infty}\!\!\ d s\ e^{-s^2}
g\left({\frac{\epsilon}{\Delta}}+s \sqrt{\frac{2{q}_0}{\Delta^2}} \right) \,\,,
\eea
where  

\begin{equation}
\label{pot}
V(w,z)=\hat{\Delta}\sigma^2 w^2 -{\lambda} w -z w\sigma\sqrt{-2{\hat{q}}_0} + \eta^{+}\theta(w)w-\eta^{-}\theta (-w)w
\end{equation}

 and the double average $\langle \dots \rangle_{\sigma,z}$ means

\be\label{DoubleAverage}
\int\limits_0^\infty\!\! \dd\sigma\ \frac{1}{N} \sum_i \delta(\sigma-\sigma_i)\int\limits_{-\infty}^\infty\!\!\frac{\dd z}{\sqrt{2\pi}}e^{-z^2/2}\ldots
\ee

Finally, the function $g$ in the integral in (\ref{free_energy}) is defined as
\begin{equation}
    g(x)= \left\{ \begin{array}{cc} 0 ,&    x\ge 0\\
    x^2 , &  -1\le x\le 0\\
    -2 x-1, & x<-1
     \end{array} \right..
\end{equation}

The differences with respect to the setup in \cite{Caccioli2018Portfolio} are the following: a trivial change of notation ($\tau$ there is $1/r$ here); the variable $\sigma$ has been introduced in (\ref{pot}), which together with the recipe (\ref{DoubleAverage}) allows us to consider assets with different volatilities $\sigma_i$ ; and the regularizer has been built into the effective potential (\ref{pot}). Note that the  $\ell_1$ in (\ref{pot}) is asymmetric, in order to allow us to penalize long and short positions separately. The usual $\ell_1$ corresponds to $\eta^{+} = \eta^{-}$, the ban on short selling to $\eta^{-}\to\infty$. We will also use the arrangement where there is a finite penalty $\eta^{-}$ on short positions and none on long ones $\eta^{+}=0$.

A note on signs: for consistency, the order parameters $\lambda$, $\Delta$, ${q}_0$ and $\hat{\Delta}$ must be positive,        ${\hat{q}}_0$ negative, and $\epsilon$ can be of either sign. Furthermore, $\lambda$ must be larger or equal to the right slope of the regularizer: $\lambda\ge\eta^{+}$.

Before setting out to derive the stationarity conditions that determine the optimal value of the free energy and thence of ES, we spell out the meaning of the order parameters. The first of these is the Lagrange multiplyer $\lambda$ that enforces the budget constraint

\be\label{eqBudgetConstraint}
\sum_{i=1}^N w_i = N.
\ee

Note that the sum of portfolio weights is set to $N$ here, instead of the usual 1. This is to keep the weights to remain of order unity in the large $N$ limit. 

Because of the relationship between $\lambda$ and the budget constraint, $\lambda$ can be thought of as a kind of chemical potential. It is an important quantity, because, as we shall see later, its value at the stationary point is equal to the free energy, hence directly related to the optimal value of ES. In \cite{Caccioli2018Portfolio} we argued that this optimal value of ES is, in fact, the in-sample estimate of Expected Shortfall. According to (\ref{equationESCost}) ES is proportional to the product $fr$, which means $f$, and hence $\lambda$ too, must be inversely proportional to $r$ when   $r=N/T\to 0$, because ES is certainly finite in this limit: a finite $N$ and $T\to\infty$ corresponds to the case of having complete information. This spurious divergence of $f$ and $\lambda$ is an artefact, due to our having absorbed a factor $1/r$ in their definition. This is explained purely by convenience: we wish to keep as close to the convention in \cite{Caccioli2018Portfolio} as possible. The opposite limit, when 
$\lambda-\eta^{+}$ vanishes, is another important point: it signals the instability of the portfolio, and the onset of the phase transition.

The next order parameter, $\epsilon$, was suggested by \cite{Rockafellar2000Optimization} as a proxy for Value at Risk. Indeed, in the limit $r\to 0$ where we know the true distribution of returns, $\epsilon$ will be seen to be equal to the known value of VaR for a Gaussian. 

The third order parameter, ${q}_0$, is of central importance: According to \cite{Caccioli2018Portfolio}, the ratio of the out-of-sample estimate ${\rm ES}_{out}$ and its true value ${\rm ES}^{(0)}$ is given by the square root of $q_0$. For the case of different $\sigma_i$'s considered here $q_0$ has to be amended by a factor depending on the structure of the portfolio\cite{Kondor2019Variancewithl1} as 

\be
\label{eqd:tildeQ}
\tilde q_0 = q_0\frac{1}{N}\sum_i\frac{1}{\sigma_i^2}.
\ee

Then the ratio of the estimated and true ES will be

\be
\label{outofsampleES}
  \frac{{\rm ES}_{out}}{{\rm ES}^{(0)}} = \sqrt{\tilde q_0}
\ee

that is the relative estimation error is $\sqrt{\tilde q_0}-1$.

The fourth order parameter, $\Delta$, can be regarded as a kind of susceptibility, it measures the reaction to a small shift in the returns.

The remaining two order parameters, ${\hat{q}}_0$ and $\hat{\Delta}$, are auxiliary variables that do not have an obvious meaning, they enter the picture through the replica formalism, and can be eliminated once the stationarity conditions have been established. The stationarity or saddle point conditions are derived by taking the derivative of the free energy with respect to the order parameters and setting them to zero. They will be written up in the next Section.

\section{Results}

First, we are going to spell out the saddle point conditions in full detail and reduce them to special cases later.

Let us bring the integral in \eqref{free_energy} to a more convenient form by integrating by parts:

\be
\label{integral}
I = \frac{1}{\sqrt{\pi}}\int\limits_{-\infty}^\infty\!\! \dd s\ e^{-s^2} g\left(\frac{\epsilon}{\Delta}+s\sqrt{\frac{2 q_0}{\Delta^2}}\right)
=\frac{2q_0}{\Delta^2}\left[W\left(\frac{\Delta+\epsilon}{\sqrt{q_0}}\right)-W\left(\frac{\epsilon}{\sqrt{q_0}}\right)\right] -1-2\frac{\epsilon}{\Delta}\,.
\ee
With this identity the free energy becomes

\be
\label{freeenergy1}
f=\lambda-\frac{\alpha\epsilon}{r}-\Delta\hat q_0-\hat\Delta q_0-\frac{\Delta}{2r}+\frac{q_0}{r\Delta}\left[W\left(\frac{\Delta+\epsilon}{\sqrt{q_0}}\right)-W\left(\frac{\epsilon}{\sqrt{q_0}}\right)\right]+\langle{\rm min} V\rangle_{\sigma,z}\,.
\ee
The function $W$ in the above formulae will frequently appear in the following, together with two related functions; they are integrals of the Gaussian $\frac{1}{\sqrt{2\pi}}e^{-x^2/2}$:

\bea
\label{Gauss_integrals}
\Phi(x) &=& \int_{-\infty}^x\mkern-18mu \dd t\ \frac{1}{\sqrt{2\pi}}e^{-t^2/2}\\
\Psi(x) &=& \int_{-\infty}^x\mkern-18mu \dd t\ \Phi(t)\\
W(x) &=& \int_{-\infty}^x\mkern-18mu \dd t\ \Psi(t) \,.
\eea
Now we evaluate the minimum of V in \eqref{pot} and denote the ``representative weight'' where this minimum is located by $w^*$. It works out to be

\be
\label{representative_weight}
w^*=\frac{\lambda+\sigma z \sqrt{-2\hat q_0}-\eta^+\Theta(w^*)+\eta^-\Theta(-w^*)}{2\sigma^2\hat\Delta}\,,
\ee
or

\be
w^*= \begin{cases}
\frac{\lambda+\sigma z \sqrt{-2\hat q_0}-\eta^+}{2\sigma^2\hat\Delta},~{\rm if}~z\ge\frac{\eta^+-\lambda}{\sigma\sqrt{-2\hat q_0}}\\
~\\
0, ~{\rm if}~-\frac{\lambda+\eta^-}{\sigma\sqrt{-2\hat q_0}}<z<\frac{\eta^+-\lambda}{\sigma\sqrt{-2\hat q_0}}\\
~\\
\frac{\lambda+\sigma z \sqrt{-2\hat q_0}+\eta^-}{2\sigma^2\hat\Delta},~{\rm if}~z\le-\frac{\lambda+\eta^-}{\sigma\sqrt{-2\hat q_0}} \,.
\end{cases}
\ee
With this and \eqref{DoubleAverage}  one can calculate $V^*$, the value of $V$ at the minimum, and perform the double averaging to obtain

\be\label{minimalpotential}
\langle V^*\rangle_{\sigma, z} = \frac{\hat q_0}{\hat\Delta}\frac{1}{N}\sum_i\left[W\left(\frac{\lambda-\eta^+}{\sigma_i\sqrt{-2\hat q_0}}\right)+W\left(-\frac{\lambda+\eta^-}{\sigma_i\sqrt{-2\hat q_0}}\right)\right].
\ee
Then the fully explicit form of the free energy becomes

\bea\label{full_freeenergy}
f&=&\lambda-\frac{\alpha\epsilon}{r}-\Delta\hat q_0-\hat\Delta q_0-\frac{\Delta}{2r}+\frac{q_0}{r\Delta}\left[W\left(\frac{\Delta+\epsilon}{\sqrt{q_0}}\right)-W\left(\frac{\epsilon}{\sqrt{q_0}}\right)\right]\\
\nonumber &+& \frac{\hat q_0}{\hat\Delta}\frac{1}{N}\sum_i\left[W\left(\frac{\lambda-\eta^+}{\sigma_i\sqrt{-2\hat q_0}}\right)+W\left(-\frac{\lambda+\eta^-}{\sigma_i\sqrt{-2\hat q_0}}\right)\right].
\eea
It is now straightforward to take the derivatives of $f$ with respect to the order parameters and derive the stationary conditions.

From $\partial f/\partial \lambda=0$ it follows that

\be\label{lambdaeq}
1=\frac{\sqrt{-2\hat q_0}}{2\hat\Delta}\frac{1}{N}\sum_i\frac{1}{\sigma_i}\left[\Psi\left(\frac{\lambda-\eta^+}{\sigma_i\sqrt{-2\hat q_0}}\right)-\Psi\left(-\frac{\lambda+\eta^-}{\sigma_i\sqrt{-2\hat q_0}}\right)\right].
\ee
The derivative with respect to $\hat q_0$ yields

\be\label{Deltaeq}
2\Delta\hat\Delta = \frac{1}{N}\sum_i\left[\Phi\left(\frac{\lambda-\eta^+}{\sigma_i\sqrt{-2\hat q_0}}\right)+\Phi\left(-\frac{\lambda+\eta^-}{\sigma_i\sqrt{-2\hat q_0}}\right)\right].
\ee
From the derivative with respect to $\hat\Delta$ we get

\be
\label{qeq}
q_0=-\frac{\hat q_0}{\hat\Delta^2}\frac{1}{N}\sum_i\left[W\left(\frac{\lambda-\eta^+}{\sigma_i\sqrt{-2\hat q_0}}\right)+W\left(-\frac{\lambda+\eta^-}{\sigma_i\sqrt{-2\hat q_0}}\right)\right].
\ee

As mentioned before, $q_0$ determines the out-of-sample estimate for ES and the estimation error.

The derivative with respect to $q_0$ leads to

\be
\label{effective_r}
2r\Delta\hat\Delta=\Phi\left(\frac{\Delta+\epsilon}{\sqrt{q_0}}\right)- \Phi\left(\frac{\epsilon}{\sqrt{q_0}}\right),
\ee
where use has been made of the identity

\be
\label{Widentity}
W(x)=\frac{1}{2}x\Psi(x)+\frac{1}{2}\Phi(x)\,.
\ee
The condition for the derivative with respect to $\epsilon$ to vanish is

\be
\label{alphaeq}
\alpha=\frac{\sqrt{q_0}}{\Delta}\left[\Psi\left(\frac{\Delta+\epsilon}{\sqrt{q_0}}\right)-\Psi\left(\frac{\epsilon}{\sqrt{q_0}}\right)\right].
\ee
The derivation of the last equation takes a little more effort. Let us go back to the free energy in \eqref{free_energy} and take the derivative with respect to $\Delta$. Noticing that $\langle V\rangle_{\sigma, z} $ does not depend on $\Delta$, and using the integral given in \eqref{integral} we have

\be
\frac{\partial f}{\partial\Delta} = -\hat q_0 +\frac{1}{2r}I+\frac{\Delta}{2r}\frac{\partial I}{\partial\Delta}=0
\ee
valid at the stationary point. From here we find

\be
\frac{1}{2r}I_{st} = \hat q_0 +\frac{2 q_0}{r\Delta^2}\left[W\left(\frac{\Delta+\epsilon}{\sqrt{q_0}}\right)-W\left(\frac{\epsilon}{\sqrt{q_0}}\right)\right]-\frac{\epsilon}{r\Delta}-\frac{\sqrt{q_0}}{r\Delta}\Psi\left(\frac{\Delta+\epsilon}{\sqrt{q_0}}\right),
\ee
where \eqref{integral} was used again. Now we apply the identity \eqref{Widentity} and the stationary conditions \eqref{alphaeq}, \eqref{effective_r} to arrive at

\be
\frac{1}{2r}I_{st} = \hat q_0 +\frac{2q_0\hat\Delta}{\Delta}-(1-\alpha)\frac{\epsilon}{r\Delta}\,,
\ee
which combined with \eqref{integral} finally leads to

\be
\label{hatqeq}
\hat q_0+\frac{2 q_0\hat\Delta}{\Delta}+\alpha\frac{\epsilon}{r\Delta}+\frac{1}{2r}-\frac{q_0}{r\Delta^2}\left[W\left(\frac{\Delta+\epsilon}{\sqrt{q_0}}\right)-W\left(\frac{\epsilon}{\sqrt{q_0}}\right)\right]=0\,.
\ee
The equations \eqref{lambdaeq}, \eqref{Deltaeq}, \eqref{qeq}, \eqref{effective_r}, \eqref{alphaeq} and \eqref{hatqeq} constitute the system of equations for the six order parameters. These equations are valid both for the regularized and (setting $\eta^+=\eta^-=0$) for the unregularized cases.

Let us now work out the relationship between the free energy and the chemical potential. Comparing \eqref{minimalpotential} and \eqref{qeq} we see that $\langle V^*\rangle_{\sigma,z}=-q_0\hat\Delta$, which with \eqref{freeenergy1} and \eqref{hatqeq} results in the simple formula

\be
\label{freeenergylambda}
f=\lambda
\ee
at the stationary point, as we anticipated before. In \cite{Caccioli2018Portfolio} we argued that the stationary value of $f$ determines the in-sample estimate of ES through \eqref{equationESCost}.

The last object to determine is the distribution of weights

\be
\label{weigthdistr}
p(w)=\langle\delta(w-w^*)\rangle_{\sigma,z}\,.
\ee

With \eqref{representative_weight} we find

\bea
\label{weightdistri1}
p(w)&=&n_0\delta(w)+\frac{1}{N}\sum_i\frac{1}{\sigma_w^{(i)}\sqrt{2\pi}}{\rm exp}\left(-\frac{1}{2}\left(\frac{w-w_i^+}{\sigma_w^{(i)}}\right)^2\right)\theta(w)\\
&+&\frac{1}{N}\sum_i\frac{1}{\sigma_w^{(i)}\sqrt{2\pi}}{\rm exp}\left(-\frac{1}{2}\left(\frac{w-w_i^-}{\sigma_w^{(i)}}\right)^2\right)\theta(-w)\,,
\eea

where

\be
\label{indvariance}
\sigma_w^{i}=\frac{\sqrt{-2\hat q_0}}{2\hat\Delta\sigma_i}
\ee

is the (estimated) variance of the $i$th return,

\be
\label{centerofpositive}
w_i^+=\frac{\lambda-\eta^+}{2\sigma_i^2\hat\Delta}
\ee
is the center of the Gaussian distribution of the (estimated) positive weight $i$,

\be
\label{centerofnegative}
w_i^-=\frac{\lambda+\eta^-}{2\sigma_i^2\hat\Delta}
\ee
is the same for negative weight $i$, and finally

\be
\label{n_0}
n_0= \frac{1}{N}\sum_i\left[\Phi\left(\frac{\lambda+\eta^-}{\sigma_i\sqrt{-2\hat q_0}}\right)-\Phi\left(\frac{\lambda-\eta^+}{\sigma_i\sqrt{-2\hat q_0}}\right)\right]
\ee
is the density of the assets whose weights are set to zero by the regularizer ($\delta(w)$ is the Dirac delta).

We wish to make and important remark here: the right hand side of \eqref{Deltaeq} is just $1-n_0$. This will prove to be the key to the mapping between the regularized and unregularized cases.

Let us record the condensate density $n_0$ also for the special case when short positions are excluded ($\eta^-\to\infty$), but long positions are not penalized ($\eta^+=0$):

\be
\label{n_0_noshort}
n_0= \frac{1}{N}\sum_i\left[ 1 -\Phi\left(\frac{\lambda}{\sigma_i\sqrt{-2\hat q_0}}\right)\right]\,.
\ee

From \eqref{n_0_noshort} we can see that, since $\Phi(x)$ is monotonic increasing and, for $x\ge 0$, concave, the contribution to $n_0$ from assets with larger $\sigma_i$'s is larger than that from smaller $\sigma_i$'s. That means that in the no-short limit the regularizer $\ell_1$ eliminates more volatile assets with larger probability than the less volatile ones. Thus we can think of the no-short constraint as an upper cutoff in volatility. This is not true in the generic case \eqref{n_0}, where the contributions of the small and large volatility items depend on the order parameters and the regularizer's slopes $\eta^+$ and $\eta^-$ in a complicated manner: the probability of an asset with volatility $\sigma_i$ to be removed is given by the difference of the two term in \eqref{n_0} under the sum. We do not wish to analyze this situation in detail, apart from the remark that a sufficiently large $\eta^-$ generally favours the elimination of large volatility items.

The integral of $p(w)$ is, of course, $1$. Its first moment, $\langle w^*\rangle_{\sigma,z}$, works out to be the same as \eqref{lambdaeq}:
\be
\label{1stmoment}
\langle w^*\rangle_{\sigma,z}=1\,.
\ee

The second moment of the weight distribution is readily obtained as

\be
\label{2ndmoment}
\langle {\left(w^*\right)}^2\rangle_{\sigma,z} = -\frac{\hat q_0}{\hat\Delta^2}\frac{1}{N}\sum_i\frac{1}{\sigma_i^2}\left[W\left(\frac{\lambda-\eta^+}{\sigma_i\sqrt{-2\hat q_0}}\right)+W\left(-\frac{\lambda+\eta^-}{\sigma_i\sqrt{-2\hat q_0}}\right)\right].
\ee

The variance of the weight distribution is then

\be
\langle {\left(w^*\right)}^2\rangle_{\sigma,z}-\left(\langle w^*\rangle_{\sigma,z}\right)^2,
\ee
which is equal to $q_0-1$, when the variances of the assets are all equal to $1$. For a portfolio with different $\sigma_i$'s, however, the relevant quantity that determines the out-of-sample estimate of ES is not the second moment of the weight distribution, but the true variance of the $i$th asset multiplied by the estimated portfolio weights squared and summed over the different assets, that is

\be
\label{q_0}
\langle {\sigma^2\left(w^*\right)}^2\rangle_{\sigma,z} \,,
\ee

which is precisely $q_0$ as given in (\ref{qeq}), and this is the quantity (multiplied by the correction as in \eqref{eqd:tildeQ}) that enters the formula for the out-of-sample estimate of ES in \eqref{outofsampleES}. For a not too inhomogeneous portfolio the difference between the second moment of the weight distribution and $q_0$ is not significant, so we can think of $q_0$ as a measure of the variance of the portfolio.

Now we are ready to consider various special cases.

\subsection{The limit of complete information}

When we have very many observations (very long time series, $T\to\infty$) relative to the dimension $N$ of the portfolio, we are in the $r=N/T\to 0$ limit. As we have already mentioned, this also corresponds to the ``chemical potential'' $\lambda$ going to infinity. Obviously, in this limit the regularizer plays no role.

We need the asymptotic behavior of the functions appearing in our stationary conditions: for $x\to\infty$, $\Phi(x)\to 1$, $\Psi(x)\sim x$ and $W(x)\sim x^2/2$, while for $x\to -\infty$ all three vanish exponentially.

Then from \eqref{lambdaeq} we have

\be
1=\frac{\lambda}{2\hat\Delta}\frac{1}{N}\sum_i \frac{1}{\sigma_i^2}\,.
\ee
From \eqref{Deltaeq}

\be
2\Delta\hat\Delta=1\,.
\ee
Combining the two:

\be
1=\lambda\Delta\frac{1}{N}\sum_i\frac{1}{\sigma_i^2}\,.
\ee
We know from \eqref{equationESCost} and \eqref{freeenergylambda} that $\lambda$ must be inversely proportional to $r$ when $r\to 0$. It follows that $\Delta\sim r$ for small $r$.

Then from \eqref{qeq} we find

\be
q_0=\Delta^2\lambda^2\frac{1}{N}\sum_i\frac{1}{\sigma_i^2}\,.
\ee
Combined with the previous equation this gives

\be
q_0 = \frac{1}{\frac{1}{N}\sum_i\sigma_i^2}\,.
\ee
The ``true'' ($r\to 0$) value of the order parameter $q_0$ is thus determined by the structural constant $\frac{1}{N}\sum_i\frac{1}{\sigma_i^2}$, which is given by the variances of the returns $\sigma_i^2$. This is in accord with the corresponding result found in the case of the  $\ell_1$-regularized variance risk measure \cite{Varga2016Replica}, \cite{Kondor2019Variancewithl1}. The above result for $q_0$ also means that the quantity $\tilde q_0$ introduced in \eqref{eqd:tildeQ} is equal to 1, and according to \eqref{outofsampleES} the out-of-sample estimate of ES is equal to its true value $\rm ES^{(0)}$, the estimation error is zero - an obvious result for the case of complete information.

From \eqref{alphaeq} with $\Delta\to 0$ we obtain $\alpha=\Phi(\epsilon/\sqrt{q_0})$, or

\be
\label{VaR}
\epsilon = \Phi^{-1}(\alpha)\sqrt{q_0}\,\,.
\ee
Now from \eqref{effective_r} we get $r=\Phi ' \left(\frac{\epsilon}{\sqrt{q_0}}\right)\frac{\Delta}{\sqrt{q_0}}$, or

\be
\Delta=r \sqrt{q_0} \frac{1}{\frac{1}{\sqrt{2\pi}}e^{-\epsilon^2/2q_0}}\,.
\ee
But then we have found

\be
\lambda=\frac{q_0}{\Delta}=\frac{1}{r}\frac{1}{\sqrt{2\pi}}e^{-\epsilon^2/2q_0}\sqrt{q_0}=\frac{1}{r}\frac{1}{\sqrt{2\pi}}e^{-\left(\Phi^{-1}(\alpha)\right)^2/2}\sqrt{q_0} \, \,.
\ee
Since $\lambda=f$ and ${\rm ES}=fr/(1-\alpha)$, we have the $r\to 0$ limit (the true value) of ES:

\be
\label{trueES}
{\rm ES}^{(0)}=\frac{1}{1-\alpha}\frac{1}{\sqrt{2\pi}}e^{-\left(\Phi^{-1}(\alpha)\right)^2/2}\sqrt{q_0}\,\,.
\ee
We record the $r\to 0$ limits of the two auxiliary variables, $\hat\Delta$ and $\hat q$, for completeness:

\be
\hat\Delta=\frac{1}{2r\sqrt{q_0}}\frac{1}{\sqrt{2\pi}}e^{-\epsilon^2/2q_0}
\ee
and
\be
\hat q_0\sim -\frac{1}{r}\,,
\ee
with a coefficient that will not be needed in the following.

Let us turn to the distribution of weights now.

In the $r\to 0$ limit the widths of the Gaussians in~\eqref{weightdistri1} all vanish, so the Gaussians become delta functions:

\be
p=\frac{1}{N}\sum_i \delta(w-w_i^+)\theta(w)+\frac{1}{N}\sum_i\delta(w-w_i^-)\theta(-w)\,.
\ee
In the $r\to 0$ limit the weights are all positive, so the second sum disappears.

For the weights $w_i^+$ we find

\be
w_i^+\simeq\frac{\lambda}{2\sigma_i^2\hat\Delta} = \frac{\lambda\Delta}{\sigma_i^2}=\frac{1}{\sigma_i^2}\frac{1}{\frac{1}{N}\sum_k \frac{1}{\sigma_k^2}}\,.
\ee
They sum to $N$, as stipulated.

The variance of a linear combination of independent random variables with averages $w_i^+$ and variances $\sigma_i^2$ is

\be
\sigma_p^2 =\sum_i\left(w_i^+\right)^2\sigma_i^2 =\frac{N}{\frac{1}{N}\sum_k\frac{1}{\sigma_k^2}}\,.
\ee

Now we recognize the meaning of the (true value of the) order parameter $q_0$: it is the normalized (to $\mathcal{O}(1)$) variance of the portfolio. This also explains the correction factor appearing in (\eqref{eqd:tildeQ}). We also see that \eqref{VaR} and \eqref{trueES} are the standard expressions for Value at Risk and Expected Shortfall indeed.

We emphasize again that all the results presented in this subsection are only valid in the $r\to 0$ limit when we are dealing with a finite dimension $N$ and infinitely long time series $T$. 

For finite $r$ the sample fluctuations start to broaden the delta spikes in the distribution of weights, the condensation of zero weights begins, $\lambda$ decreases and all the formulae above become considerably more complicated. We turn to this situation in the next subsections. 

By now we have learned everything that was to be learned from keeping the variances $\sigma_i$ different, in particular the elimination by the regularizer of the most volatile assets in the case of restriction of short selling.. In order to simplify the presentation and avoid the appearance of very large and hardly transparent formulae, henceforth we set all the $\sigma_i$'s equal to 1. We stress, however, that the main message of this paper, namely the existence of a mapping between the regularized and unregularized cases, depends only on the structure of the equations, and works also with different $\sigma$'s.

\subsection{Without regularization}

In this subsection we set $\eta^+=\eta^-=0$, that is we consider our problem without regularization, and according to what has just been said, put $\sigma_i=1$. We will make use of the identities

\bea
\label{Gauss_integral_identities}
\Phi(x) + \Phi(-x)&=&1\\
\Psi(x)  + \Psi(-x)&=& x\\
W(x) +W(-x)&=&\frac{1}{2}(x^2+ 1)\,.
\eea

The free energy (\eqref{full_freeenergy}) becomes

\bea\label{full_freenoreg}
f&=&\lambda-\frac{\alpha\epsilon}{r}-\Delta\hat q_0-\hat\Delta q_0-\frac{\Delta}{2r}+\frac{q_0}{r\Delta}\left[W\left(\frac{\Delta+\epsilon}{\sqrt{q_0}}\right)-W\left(\frac{\epsilon}{\sqrt{q_0}}\right)\right] - \frac{\lambda^2 }{4\hat\Delta} + \frac{\hat{q}_0}{2\hat\Delta}.                                           
\eea

For the saddle point equations we find:

\be
\label{lambdaeqnoreg}
1=\frac{\lambda}{2\hat\Delta}\,\, ,
\ee

\be
\label{Deltaeqnoreg}
2\Delta\hat\Delta = 1\,\,,
\ee

\be
\label{qeqnoreg}
q_0=\frac{\lambda^2}{4\hat\Delta^2} - \frac{\hat q_0}{2\hat\Delta^2}\,\,,
\ee

\be
\label{effective_rnoreg}
2r\Delta\hat\Delta=r= \Phi\left(\frac{\Delta+\epsilon}{\sqrt{q_0}}\right)- \Phi\left(\frac{\epsilon}{\sqrt{q_0}}\right)\,,
\ee

\be
\label{alphaeqnoreg}
\alpha = \frac{\sqrt{q_0}}{\Delta}\left[ \Psi\left(\frac{\Delta+\epsilon}{\sqrt{q_0}}\right)- \Psi\left(\frac{\epsilon}{\sqrt{q_0}}\right)\right]\,,
\ee

\be
\label{hatqeqnoreg}
\hat q_0 +\frac{2q_0\hat\Delta}{\Delta} +\frac{\alpha\epsilon}{r\Delta}+\frac{1}{2r}-\frac{q_0}{r\Delta^2}\left[ W\left(\frac{\Delta+\epsilon}{\sqrt{q_0}}\right)- W\left(\frac{\epsilon}{\sqrt{q_0}}\right)\right]=0\,.
\ee

These equations are rather similar to their counterparts in the previous subsection, but of course $r\to 0$ is not assumed here. As for their solutions, they were discussed and illustrated in several figures in \cite{Caccioli2018Portfolio}, therefore we will not dwell upon them here. (Some results will be given in Subsection 3.6.)  Instead, we write up the corresponding equations in the case where no short positions are allowed and make a term-by-term comparison between the two sets of equations.

\subsection{No short selling}

Short positions will be excluded by imposing infinite penalty on them by letting $\eta^-$ go to infinity. The functions $\Phi(x)$, $\Psi(x)$ and $W(x)$ all vanish when $x\to-\infty$. Long positions will not be penalized, so we set $\eta^+=0$.

The free energy becomes

\bea
\label{full_feenoshort}
f&=&\lambda-\frac{\alpha\epsilon}{r}-\Delta\hat q_0-\hat\Delta q_0-\frac{\Delta}{2r}+\frac{q_0}{r\Delta}\left[W\left(\frac{\Delta+\epsilon}{\sqrt{q_0}}\right)-W\left(\frac{\epsilon}{\sqrt{q_0}}\right)\right]\\
&+& \frac{\hat q_0}{\hat\Delta} W\left(\frac{\lambda}{\sqrt{-2\hat q_0}}\right).                                           
\eea

The stationary conditions now read as:

\be
\label{lambdaeqnoshort}
1=\frac{\sqrt{-2\hat q_0}}{2\hat\Delta} \Psi\left(\frac{\lambda}{\sqrt{-2\hat q_0}}\right),
\ee

\be
\label{Deltaeqnoshort}
2\Delta\hat\Delta = \Phi\left(\frac{\lambda}{\sqrt{-2\hat q_0}}\right),
\ee

\be
\label{qeqnoshort}
q_0=-\frac{\hat q_0}{\hat\Delta^2} W\left(\frac{\lambda}{\sqrt{-2\hat q_0}}\right),
\ee

\be
\label{effective_rnoshort}
2r\Delta\hat\Delta= \Phi\left(\frac{\Delta+\epsilon}{\sqrt{q_0}}\right)- \Phi\left(\frac{\epsilon}{\sqrt{q_0}}\right),
\ee

\be
\label{alphaeqnoshort}
\alpha = \frac{\sqrt{q_0}}{\Delta}\left[ \Psi\left(\frac{\Delta+\epsilon}{\sqrt{q_0}}\right)- \Psi\left(\frac{\epsilon}{\sqrt{q_0}}\right)\right],
\ee

\be
\label{hatqeqnoshort}
r\left(\hat q_0 +\frac{2q_0\hat\Delta}{\Delta}\right) +\frac{\alpha\epsilon}{\Delta}+\frac{1}{2}-\frac{q_0}{\Delta^2}\left[ W\left(\frac{\Delta+\epsilon}{\sqrt{q_0}}\right)- W\left(\frac{\epsilon}{\sqrt{q_0}}\right)\right]=0\,,
\ee
the last equation being the same as \eqref{hatqeqnoreg}, just multiplied through by r.

In the distribution of weights in \eqref{weightdistri1} the second sum of Gaussians will disappear, because for $\eta^-\to\infty$ all the weights \eqref{centerofnegative} go to infinity. The weights \eqref{centerofpositive} become 
\be
w_i^+ = \frac{\lambda}{2\hat\Delta}\,\,,
\ee
while the density of zero weights is now

\be
\label{n_0noshort}  
n_0 = 1-\Phi\left(\frac{\lambda}{\sqrt{-2\hat q_0}}\right),
\ee
which with \eqref{Deltaeqnoshort} leads to

\be
\label{n_01}
1-n_0 = 2\Delta\hat\Delta\,.
\ee

From \eqref{n_0noshort} we see that $n_0=0$ for $r=0$ and increases as $\lambda$ decreases, until it reaches its maximal value $1/2$ when $\lambda$ vanishes. Mathematically, there is nothing to prevent us from continuing to increase $r$ and driving $\lambda$ to negative values, which would allow $n_0$ to grow beyond $1/2$, up to $n_0=1$, but a negative $\lambda$ would cause the free energy and thus also ES to change sign - an extreme case of ``in-sample optimism'', entirely due to the lack of sufficient information. We consider such a situation ``unphysical'', and never go beyond the point where $\lambda$ (or $\lambda-\eta^+$ if $\eta^+>0$) vanishes anywhere in this paper.   

\subsection{No-short mapping}

We are now ready to spell out the mapping between the no-short case and the unregularized one. 

The first point to notice is that the only difference between equation \eqref{effective_rnoreg} valid in the unregularized case and its counterpart \eqref{effective_rnoshort} in the no-short case (combined with \eqref{n_01}) appears on their left hand side: the terms $r$, and $(1-n_0)r$, respectively.
This suggests to introduce an effective $r$:

\be
\label{effectiver}
r_{\rm eff} = (1-n_0)r\,. 
\ee

Now $r=N/T$, and $n_0$ is the density of the assets killed by the regularizer, thus $(1-n_0)r=\frac{N-N_0}{T}$ is the number of surviving assets divided by the length of the time series. As $r_{\rm eff}$ increases from zero to $1/2$, $r$ will increase between zero and 1.

Inspired by the connection between $r$ and $r_{\rm eff}$ we compare the two sets of equations and recognize that, in fact, the whole system of saddle point equations can be mapped from the regularized case to the unregularized one. A variable that appears in all the subsequent equations is

\be
\label{zdef}
z=\frac{\lambda}{\sqrt{-2\hat q_0}} ,
\ee
where the variables $\lambda$ and $\hat q_0$ are those that appear in the no-short equations.

Then the connection between the order parameters belonging to the two cases is the following:

\be
\label{qmap}
q_0=q_0^{\rm eff}\frac{z}{\Psi(z)}\,\,,
\ee

\be
\label{Deltamap}
\Delta=\Delta_{\rm eff} \sqrt{\frac{z}{\Psi(z)}}\,\,,
\ee

\be
\label{epsilonmap} 
\epsilon=\epsilon_{\rm eff} \sqrt{\frac{z}{\Psi(z)}}\,\,,
\ee

\be
\label{lambdamap}
\lambda=\lambda_{\rm eff}\sqrt{\frac{z}{\Psi(z)}}\Phi(z)\,,
\ee

\be
\label{hatqmap}
\hat q_0 = \hat q_0^{\rm eff}\Phi(z)\,,
\ee

\be
\label{hatDeltamap}
\hat\Delta=\hat\Delta_{\rm eff} \sqrt{\frac{\Psi(z)}{z}}\Phi(z)\,.
\ee

A direct substitution shows that if the order parameters on the left hand sides of the above equations satisfy the no-short equations, then the effective variables satisfy the unregularized ones, provided we also replace $r$ with $r_{\rm eff}$. In particular, the contour maps of the unregularized order parameters presented in \cite{Caccioli2018Portfolio} can be taken over and simply blown up by a factor $\frac{1}{1-n_0}$ to obtain the contour maps of the no-short variables. Given the relation between $q_0$ and the estimation error, we see that the mapping also means that a given error belongs to a larger $r$ in the no-short case than in the unregularized one, in other words the no-short constrained problem demands $(1-n_0)$ times less data (shorter time series) than the unregularized one. 

One may wonder whether this mapping expresses some symmetry of the problem, that is whether the free energy functional is invariant under this mapping. The answer is no: the mapping works only in the saddle point equations, it is a property of the stationary point.

It is important to learn the range of this transformation. In the limit $r\to 0$ the transformation is the identity, but this is trivial: when we have complete information the regularizer does not play any role. It is more interesting to consider the vicinity of the phase transition in the unregularized case, where $q_0^{\rm eff}$ and $\Delta_{\rm eff}$ diverge. These divergences are removed by the mapping, no singularity is found in the no-short case. This is in accord with \cite{Caccioli2016Lp}: the infinite penalty on short positions precludes the phase transition and no singularity shows up in $q_0$ , $\Delta$ or $\epsilon$. Mathematically, we can continue the unregularized solutions into the non-feasible region beyond the phase boundary, but they make no sense there (for example $q_0$ changes sign, $\Delta$ and $\epsilon$ become imaginary, etc.), while their mapped counterparts continue to behave reasonably. According to \eqref{effectiver}, when $r_{\rm eff}$ reaches the critical point $r_c(\alpha)$ the corresponding value of $r$ in the no-short problem will be twice as large, so the whole phase diagram is multiplied by a factor 2. Beyond the mapped phase boundary the regularized solutions still survive, but their meaning becomes questionable, because the free energy, hence also ES change sign. As noted in the previous subsection, we refrain from the discussion of this unphysical region.

\subsection{Mapping for generic $\ell_1$ constraint}

The mapping between the generic $\ell_1$-constrained ES optimization and the unregularized one is a straightforward generalization of the results in the previous Subsection. The mapping is made more complicated because of the sums and differences of the $\Psi$, $\Phi$ and $W$ functions appearing on the right hand side of equations \eqref{lambdaeq}, \eqref{Deltaeq} and \eqref{qeq}. We introduce the following notation for these combinations:

\be
\label{defAPsi}
A_{\Psi}= \Psi\left(\frac{\lambda-\eta^+}{\sqrt{-2\hat q_0}}\right)-\Psi\left(-\frac{\lambda+\eta^-}{\sqrt{-2\hat q_0}}\right),
\ee

\be
\label{defAPhi}
A_{\Phi}=\Phi\left(\frac{\lambda-\eta^+}{\sqrt{-2\hat q_0}}\right)+\Phi\left(-\frac{\lambda+\eta^-}{\sqrt{-2\hat q_0}}\right),
\ee

and
\be
\label{defAW}
A_W=W\left(\frac{\lambda-\eta^+}{\sqrt{-2\hat q_0}}\right)+W\left(-\frac{\lambda+\eta^-}{\sqrt{-2\hat q_0}}\right),
\ee
where we have set all the $\sigma_i =1$.

In terms of these quantities the generic map reads as

\be
\label{qmapgen}
q_0=q_0^{\rm eff}\frac{2A_W-A_{\Phi}}{(A_{\Psi})^2}\,,
\ee

\be
\label{Deltamapgen}
\Delta=\Delta_{\rm eff} \frac{\sqrt{2A_W - A_{\Phi}}}{A_{\Psi}}\,,
\ee

\be
\label{epsilonmapgen} 
\epsilon=\epsilon_{\rm eff}\frac{\sqrt{2A_W - A_{\Phi}}}{A_{\Psi}}\,,
\ee

\be
\label{lambdamapgen}
\lambda=\lambda_{\rm eff}\frac{zA_{\Phi}}{\sqrt{2A_W - A_{\Phi}}}\,,
\ee

\be
\label{hatqmapgen}
\hat q_0 = \hat q_0^{\rm eff} A_{\Phi}\,,
\ee

\be
\label{hatDeltamapgen}
\hat\Delta=\hat\Delta_{\rm eff} \frac{A_{\Phi}A_{\Psi}}{\sqrt{2A_W - A_{\Phi}}}\,.
\ee

For the condensate density $n_0$ we have

\be
\label{condensategen}
1-n_0=A_{\Phi}\,,
\ee
and for the effective aspect ratio

\be
\label{effrgen}
r_{\rm eff}=2r\Delta\hat\Delta=rA_{\Phi}=(1-n_0)r\,.
\ee

As before, if the order parameters satisfy the regularized stationarity conditions \eqref{lambdaeq} - \eqref{hatqeq} (with $\sigma_i=1$), then the effective parameters will satisfy the unregularized equations \eqref{lambdaeqnoreg} -- \eqref{hatqeqnoreg}, and vice versa.

Note that the above equations remain invariant if we redefine $\lambda$ as $\lambda-\eta^+$ and $\eta^-$ as $\eta^- +\eta^+$. So we can set $\eta^+=0$ and $\eta^- +\eta^+=\eta$ without loss of generality. We will use this setup in the following, in order to reduce the number of parameters when solving the stationarity equations.

\subsection{Solutions for the order parameters}

Except for a few exceptional points, it is impossible to obtain the solutions of the stationarity equations in closed, analytical form, but it is perfectly possible to get them numerically, by a computer. (The case of $\alpha=1$ is exceptional in several respects and will not be considered here.) In the following, the solutions will be presented in graphical form. 

\begin{figure}[htbp]
\centerline{\includegraphics[width=0.5\textwidth]{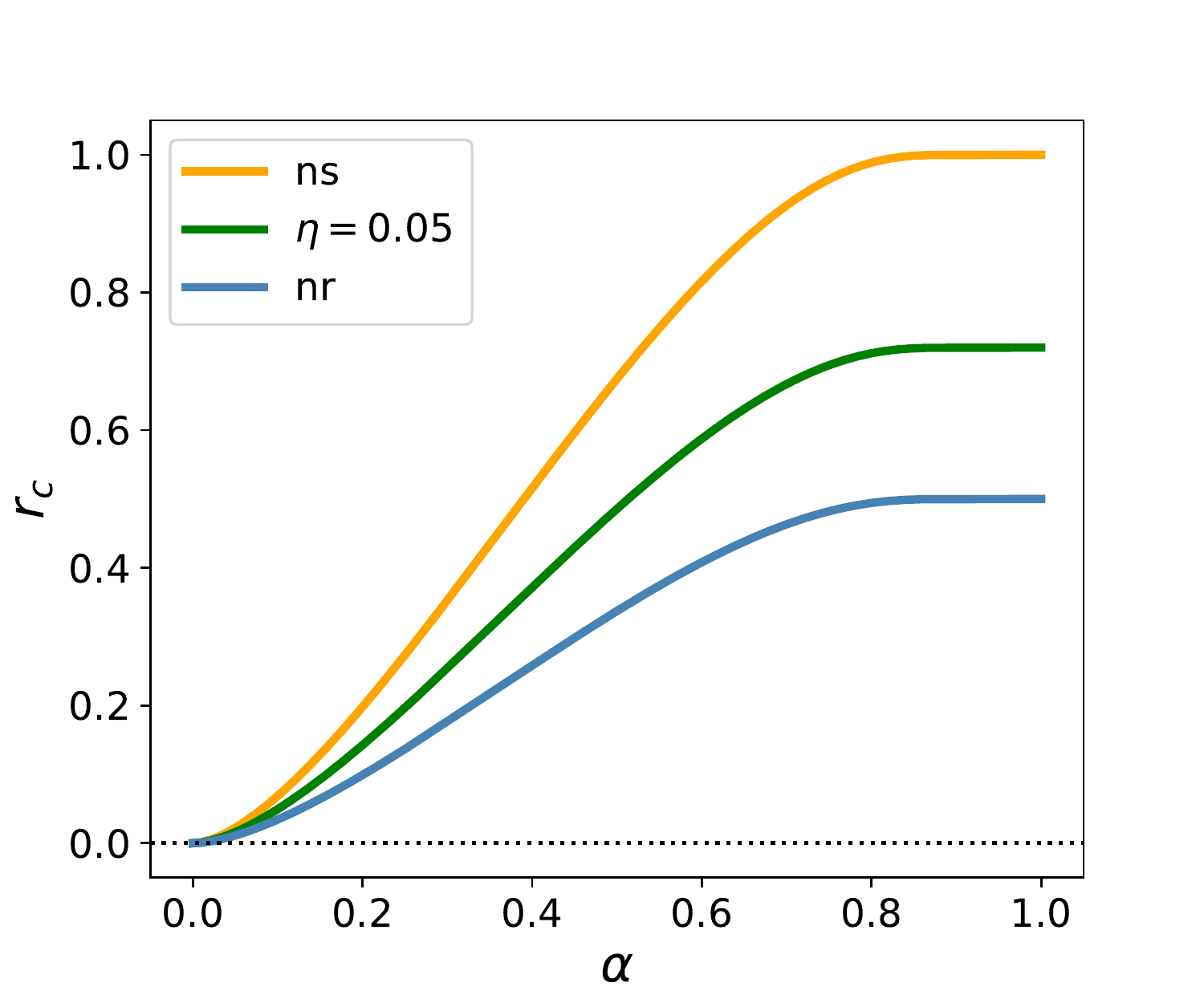}}

\caption{The boundary of the region where the optimization of ES is feasible in the unregularized case (nr); its image under the map for a finite $\eta^-=0.05$, $\eta+=0$ regularizer; and the same under the no-short map (ns). \label{fig:feas_region}}
\end{figure}

Figure~\ref{fig:feas_region} exhibits three special lines, belonging to three different cases: the unregularized case, the one with a finite regularizer, and the one with a no-short constraint.

The blue line is the upper boundary of the region where the optimization of unregularized ES is feasible. This line was first determined in~\cite{Ciliberti2007On}. It is a phase boundary, along which a phase transition takes place: $q_0$, $\Delta$ and $\epsilon$ diverge here, while $\lambda$ becomes zero. The unregularized equations can be solved also above this line, up to the horizontal line at $r=1$ (not shown in the Figure), but the solutions are meaningless: $q_0$ is negative, while $\lambda$,         $\Delta$ and $\epsilon$ become imaginary. The unregularized equations do not have any solution above $r=1$.

The green line is the image of the unregularized phase boundary under the mapping described in the previous Subsection, and corresponds to a one-sided regularizer with $\eta^-=0.05$, $\eta+=0$. There is no phase transition when we cross this line, the order parameters remain smooth, finite quantities, but $\lambda$ (along with the free energy and the in-sample estimate of ES) changes sign, rendering the solution in the region above the green line ``unphysical''.  Nevertheless, if we keep following the solutions beyond the green line we can go up to the image of the $r=1$ line (mapped into $r\to\infty$), where $q_0$ and $\Delta$ will ultimately diverge. The region between the green line and the image of the $r=1$ line has an intricate structure, but because it corresponds to negative risk, it is of no interest for us in the present context.

In the no-short case there is always a solution with the order parameters remaining finite all the way up to infinity, which is the image of the $r=1$ line under the no-short map. However, along the orange line $\lambda$ changes sign, and the region beyond it is meaningless again. The orange line is the unregularized phase boundary (blue line) blown up by a factor $\frac{1}{1-n_0}=2$. All this is in accord with the picture described in \cite{Caccioli2016Lp} in that the no-short constraint eliminates the critical line. The solutions becoming unphysical beyond a certain $r$-range could not be foreseen on the basis of the analysis in~\cite{Caccioli2016Lp}.

\begin{figure}[htbp]
\centerline{\includegraphics[width=\textwidth]{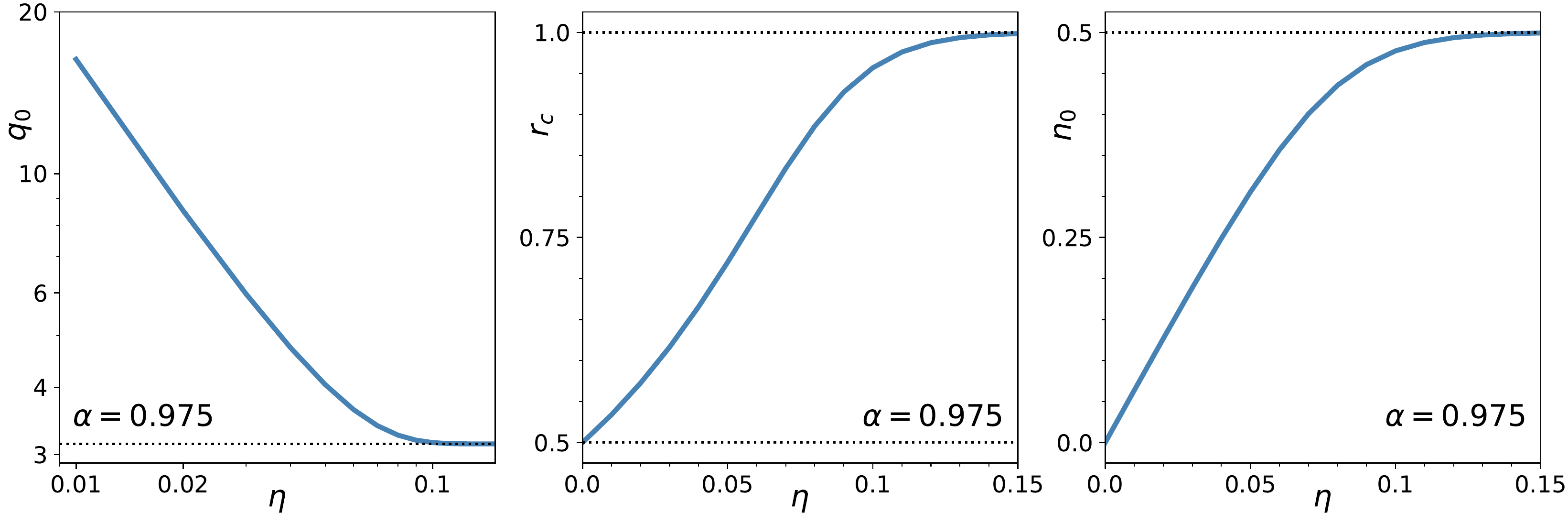}}
\caption{Dependence of $q_0$ at $r_c$ (left), critical point (middle) and proportion of zero weigths at $r_c$ (right) as a function of the regularization strength, $\eta^-=\eta$ ($\eta^+=0$). Note the logarithmic scale in the left panel. \label{fig:eta_dep}}
\end{figure}

Figure~\ref{fig:eta_dep} shows the $\eta$-dependence of $q_0$ and the density of the zero weights $n_0$ at criticality, and that of the value of the critical $r$. In the unregularized case ($\eta\to 0$), $q_0\to\infty$, while in the no-short case ($\eta\to\infty$) $q_0\to\pi$. The value of the critical $r_c$ increases from $r_c\approx1/2$ at $\alpha=0.975$ in the unregularized case to $\approx 1$ for the no-short case. The proportion of the assets eliminated from the portfolio (the condensate density) goes from zero for $\eta=0$ to $1/2$ for large $\eta$.

\begin{figure}[htbp]
\centerline{\includegraphics[width=0.32\textwidth]{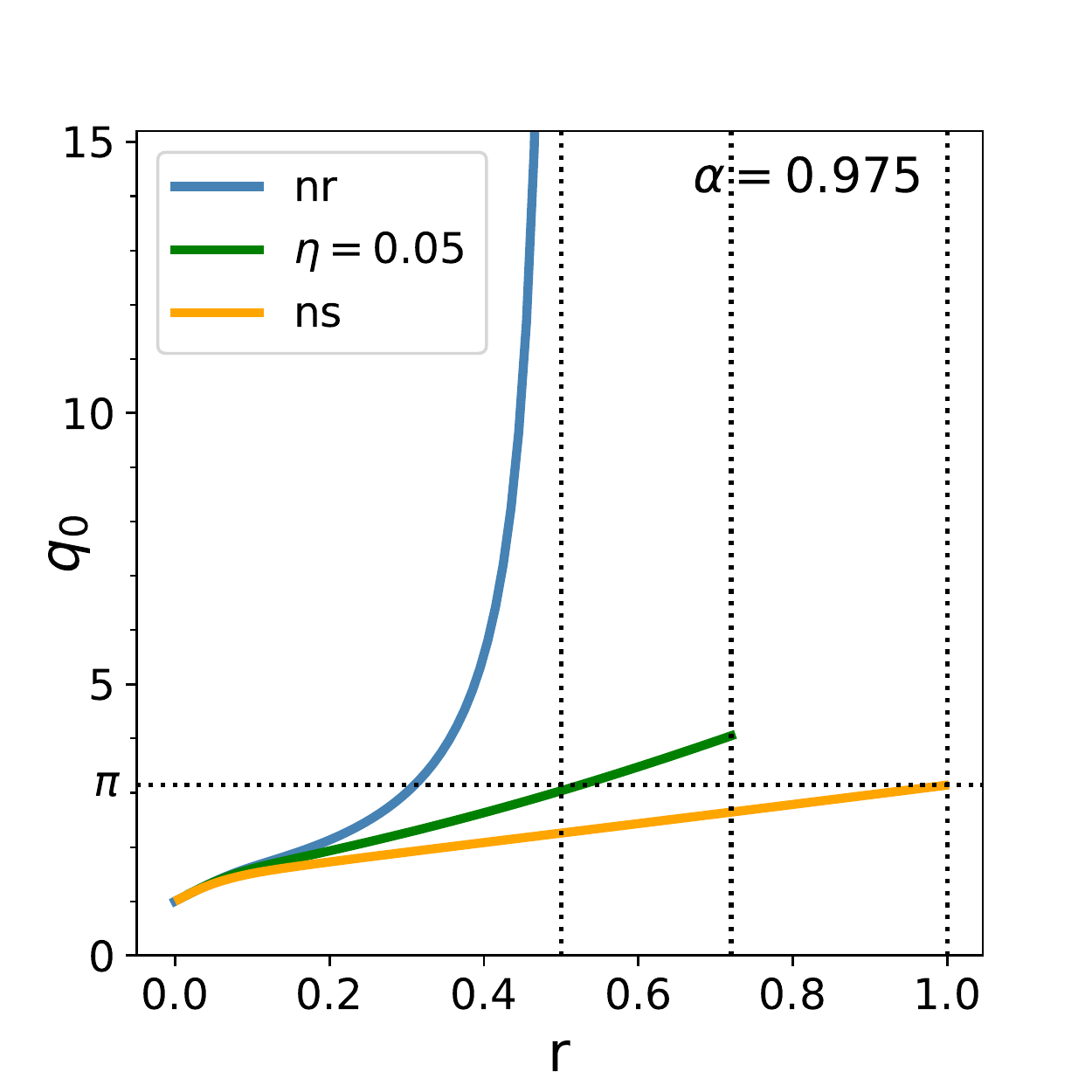} \kern1em
                 \includegraphics[width=0.32\textwidth]{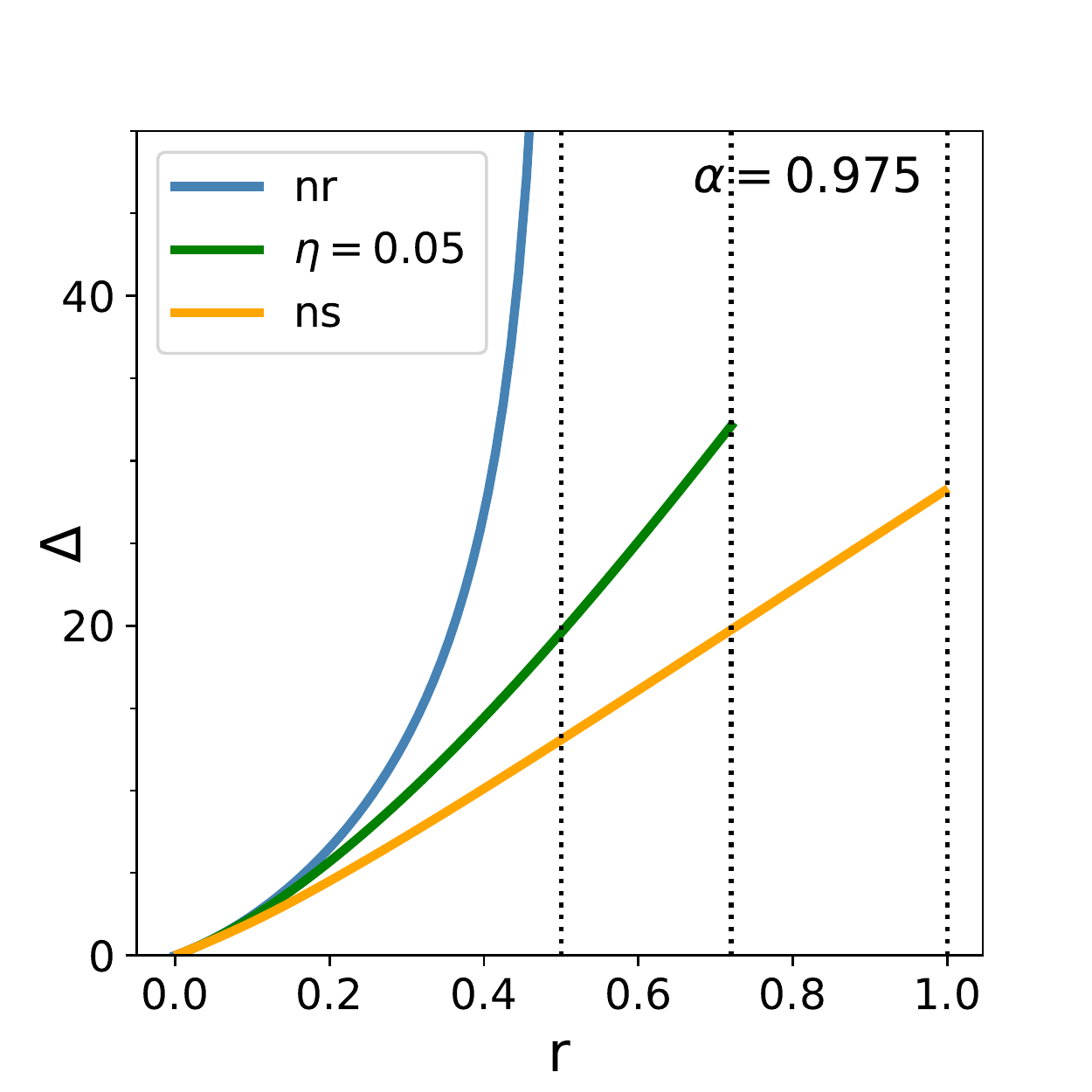} \kern1em
                 \includegraphics[width=0.32\textwidth]{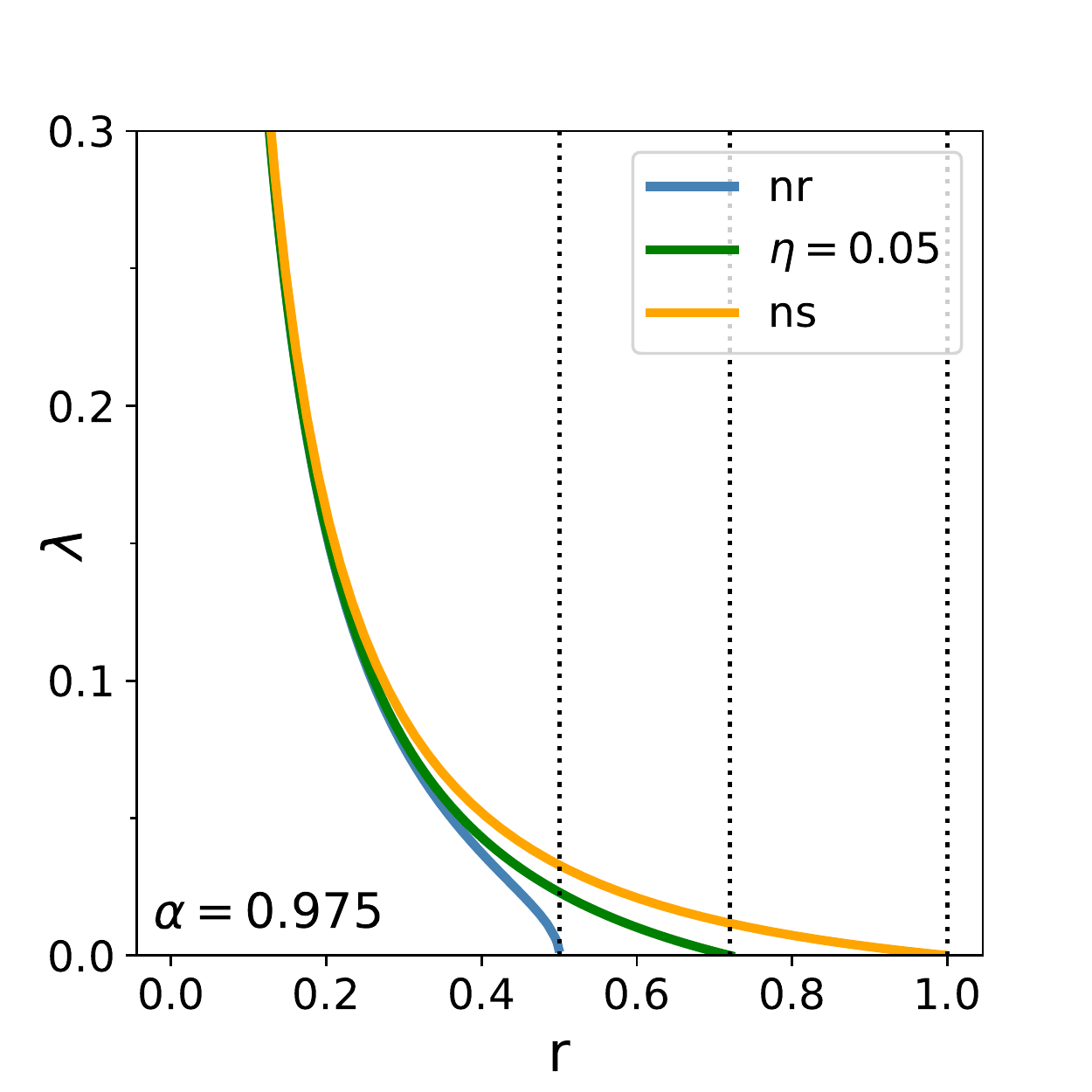}} 
\caption{Dependence of $q_0$(left), $\Delta$ (middle) and ``chemical potential'' $\lambda$ (right) on $r=N/T$, for the unregularized (blue), $\eta^-=0.05, \eta^+=0$ regularized (green), and no-short (yellow) cases.\label{fig:qDl}}
\end{figure}

In Figure~\ref{fig:qDl} we display the $r$-dependence of $q_0$, $\Delta$ and $\lambda$ for the three cases: unregularized, regularized and no-short. Without regularization $q_0$ and $\Delta$ increase with $r$ and diverge at an $r_c$ slightly less than $\frac{1}{2}$; while $\lambda$ decreases from infinity at $r=0$ to zero at $r_c$. (The confidence limit $\alpha$ is set at its regulatory value 0.975 in these figures.) Under the regularizer $\eta^-=0.05$, $\eta+=0$, $q_0$ and $\Delta$ increases up to the $r$ where $\lambda$ vanishes. The situation is similar for an infinitely strong (no-short) regularizer, with the limiting value of $q_0 = \pi$ and $\lambda =0$ at $r\approx1$.

\begin{figure}[htbp]
\centerline{\includegraphics[width=0.32\textwidth]{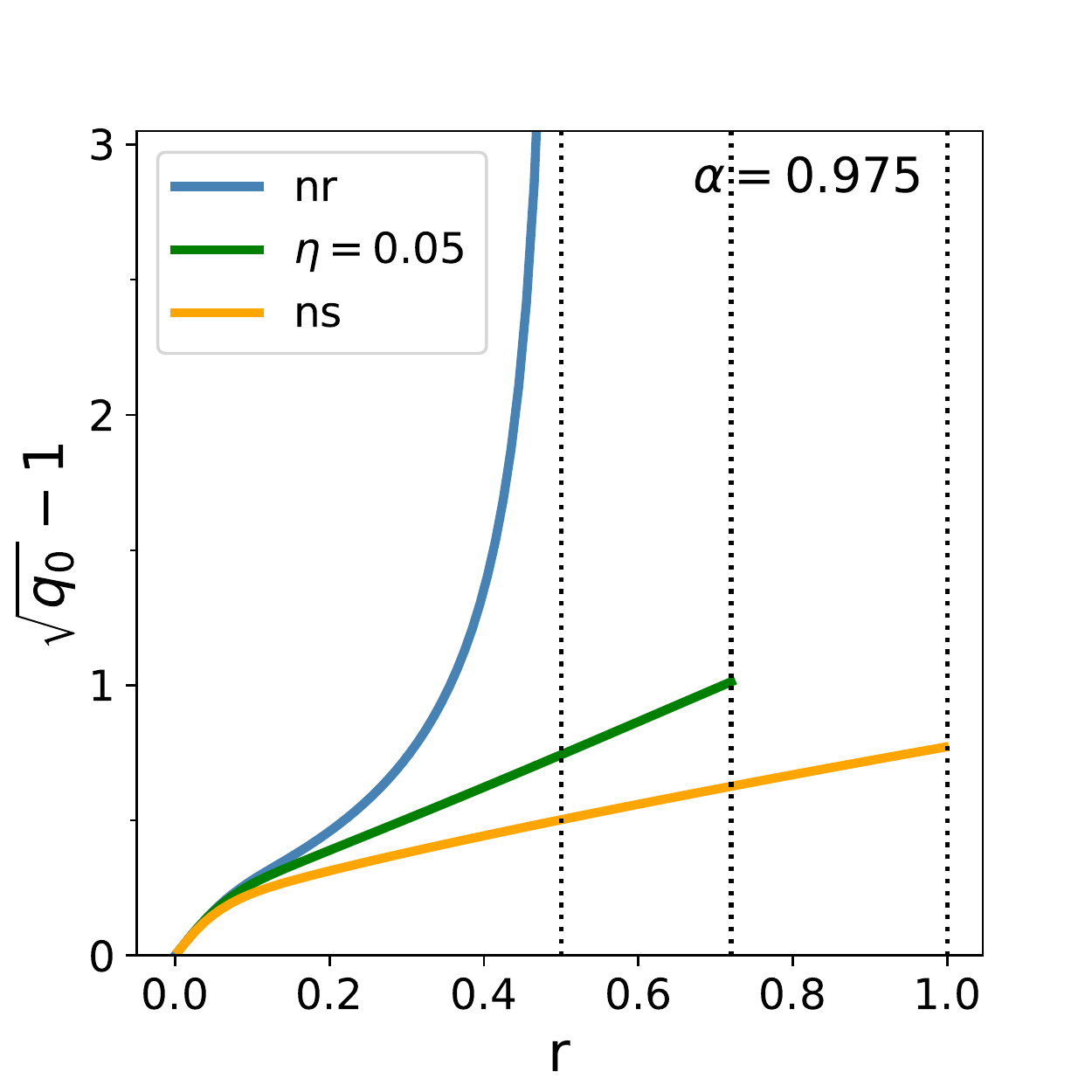} \kern1em
                 \includegraphics[width=0.32\textwidth]{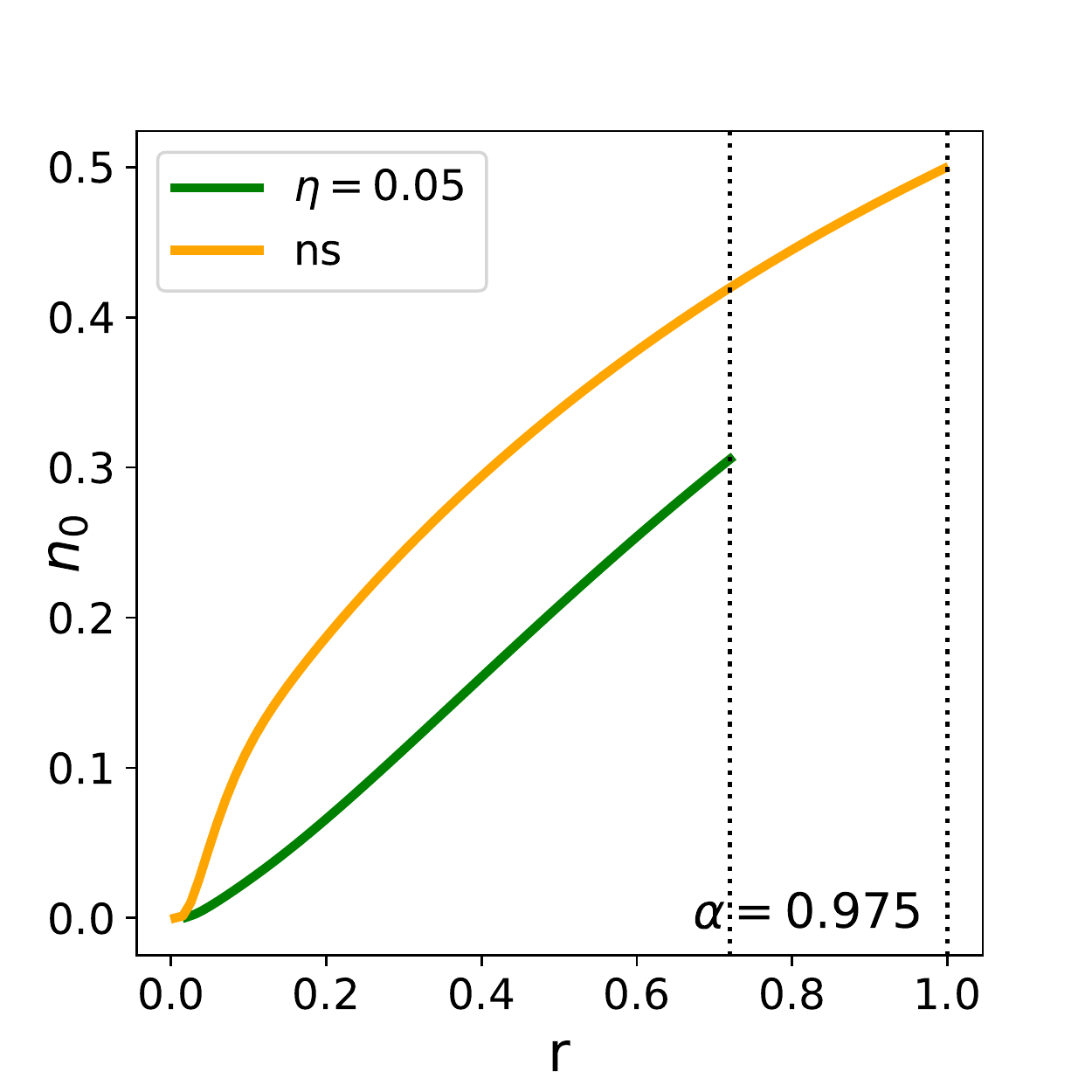} \kern1em
                 \includegraphics[width=0.32\textwidth]{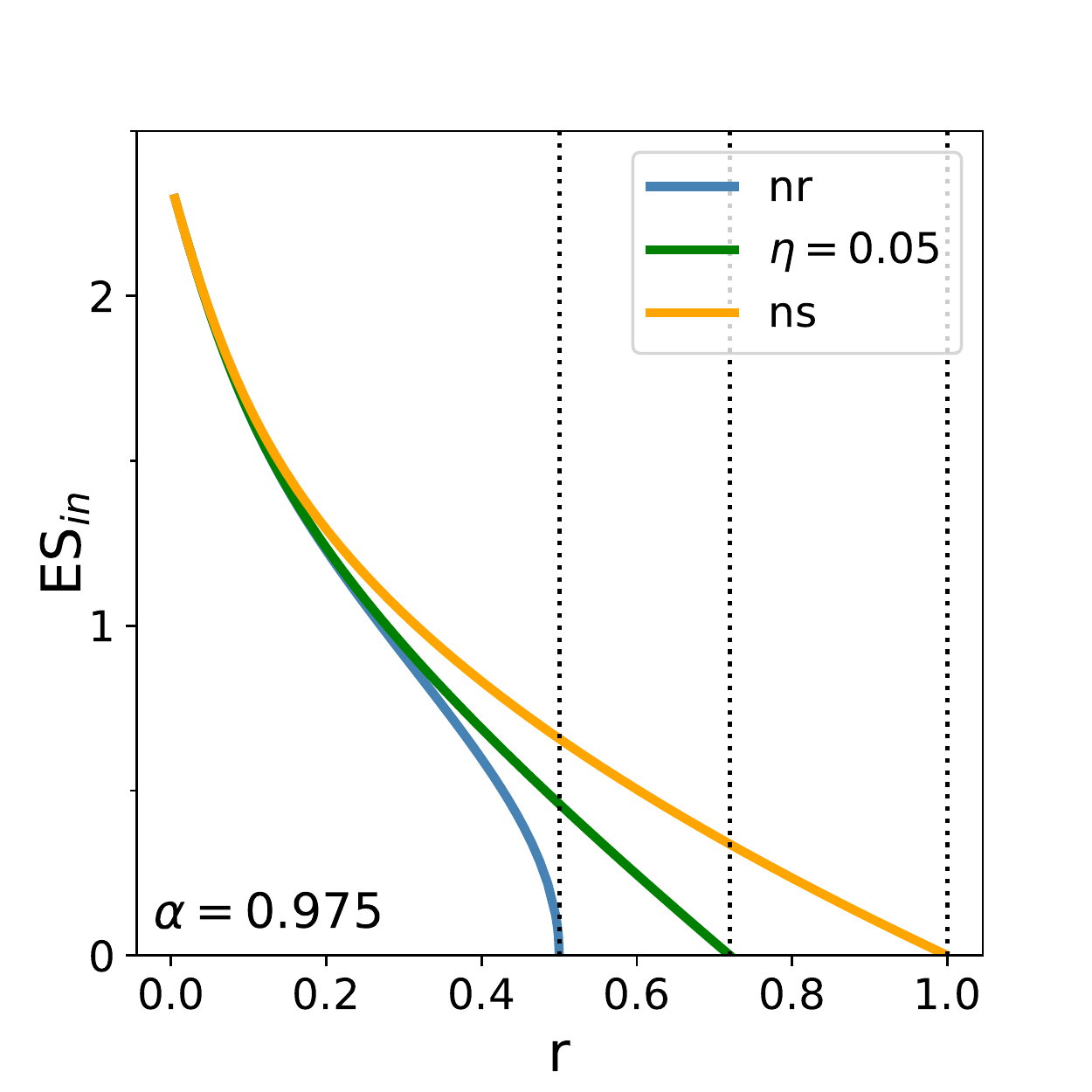}} 

\caption{Dependence of the out-of-sample estimation error (left), proportion of zero weights (center) and in-sample ES (right) on $r=N/T$, for the non-regularized (blue), $\eta^-=\eta$ ($\eta^+=0$) regularized (green), and no-short (orange) cases.\label{fig:EEnES}}
\end{figure}

The left panel in Figure~\ref{fig:EEnES} shows the relative out-of-sample estimation error, which is related to the out-of-sample estimate of ES by \eqref{outofsampleES} ($\tilde{q}_0=q_0$ now, as we have set all the $\sigma_i$ =1). These curves are similar to the curves of $q_0$ in the previous Figure. It can be seen that the curves of the relative estimation error run very close to each other for small values of $r$: there is no substantial reduction of the error in this range. Where they fan out and the effect of regularization starts to be felt (say around $r=0.1$), the relative error is already about 20 \%.

The middle panel in Figure~\ref{fig:EEnES} shows the behaviour of the density of zero weights as function of $r$ for the finite $\eta$-regularized and the no-short cases. In the no-short case $n_0$ reaches its maximal value $\frac{1}{2}$ at $r\approx1$ (for $\alpha=0.975$) where $\lambda$ vanishes. For a regularizer of finite strength it always remains below $\frac{1}{2}$. 

The right panel in Figure~\ref{fig:EEnES} displays the behaviour of the in-sample estimate of ES for the three cases. This quantity is directly related to $\lambda$ through \eqref{equationESCost} and \eqref{freeenergylambda}. The monotonic and fast decay of these curves demonstrates what is called in-sample optimism, a strong underestimation of risk.

\section{Discussion}

In the preceding Section we compared the behaviour of the order parameters in the three instances considered in this paper: the case of the unregularized, the $\ell_1$-regularized, and the no-short constrained Expected Shortfall optimization. We have seen that without regularization there is a phase transition as we cross the phase boundary $r_c(\alpha)$ shown in Figure~\ref{fig:feas_region} with $\Delta$, $q_0$ and $\epsilon$ diverging here, as known since the paper \cite{Ciliberti2007On}. In contrast, the infinite penalty on short positions suppresses this phase transition, while an $\ell_1$ regularizer with finite slopes only shifts the phase boundary. These facts were also known from earlier work \cite{Caccioli2016Lp}, \cite{Caccioli2013Optimal}. However, the picture has turned out to be more complicated than envisaged in \cite{Caccioli2016Lp}. The numerical solution for the order parameters performed in this paper has revealed that new characteristic lines emerge both in the case of finite regularization and the no-short constraint, along which the order parameter $\lambda$ and, consequently, the free energy and the in-sample estimate of Expected Shortfall change sign. We have determined the position of these new characteristic lines: in the no-short case the new line is the curve $2r_c(\alpha)$, for a finite regularizer it is $\frac{r_c(\alpha)}{1-n_0}$, where $n_0\le\frac{1}{2}$. We have omitted the detailed analysis of the regions above these lines, where the estimated risk becomes negative. Instead, we confined ourselves to merely pointing out that the critical line for the no-short constraint is projected out to infinity, so the phase transition is removed indeed, while for a finite slope regularizer the critical line is shifted into the unphysical, negative risk region, where for some values of the regularizer's strength $\eta$ it even develops two branches.

We have also found the behaviour of the various order parameters, most notably that of $q_0$ that determines the out-of-sample estimation error of ES, the free energy that gives the in-sample estimator, and the susceptibility-like quantity $\Delta$, and displayed their behaviour for the three cases studied here. It is satisfactory to see that $q_0$ and $\Delta$ remain finite up to the new characteristic lines, that is the regularizer acts as expected: it suppresses the divergent sample fluctuations in the optimization of ES. Unfortunately, this suppression is not strong enough to bring down the estimation error to acceptable values, except for the range of  small $r=\frac{N}{T}$ ratios where it demands far too long time series for any realistic $N$, and where $r$ is small already without any regularization. 

What is the meaning of this phase transition? As analyzed in \cite{Vargahaszonits2008Instabilityofdownside} and \cite{Kondor2010Instability} it follows from the coherence axioms that coherent risk measures, including ES, are unstable in the sense that whenever an asset or a combination of assets in the portfolio stochastically dominates the others in a given sample, the investor can take an extremely large long position in the dominant asset and compensate this with an appropriately large short position, without violating the budget constraint. This means that the weight of the dominant asset runs away  practically to infinity, resulting in an arbitrarily large negative value of the risk measure. This is a mirage of an arbitrage, which can disappear in the next sample, or change into another arbitrage with a different weight running away to infinity. In practice, there are always constraints that prevent such a divergence from taking place. The ban on short selling is just this sort of constraint. The runaway solutions try to escape, but get arrested at the walls constituted by the constraint, in the case of a no-short ban, at the coordinate planes. This is how the condensate of zero weights builds up. This mechanism is the stronger the larger the ratio $r=N/T$. 

There is nothing surprising about solutions sitting on the constraint-walls or at corners in a linearly programmable problem, such as the optimization of ES. In the usual applications of linear programming the constraints typically express some physical limitation like a finite amount of resources, material or labor, etc. In the present finance problem such a finite resource would be the limited budget, but if short selling is not constrained the budget in itself can not prevent runaway solutions. The ban on short positions corresponds to an infinitely strong $\ell_1$ regularizer, which, combined with the budget constraint, is already sufficient to take care of the runaway solutions. So with a no-short ban on, we can increase $r$ (that is the dimension, or decrease the amount of data) without any mathematical contradiction showing up; neither $q_0$ nor $\Delta$ will diverge. It is clear, however, that the solution based on less and less information becomes increasingly meaningless. In these circumstances the optimization will not tell us anything useful about the structure of the market, it will be determined more and more by the constraint.

What we regard as the most intriguing result of this paper is the existence of a mapping between the regularized and the unregularized problems. 

{\bf Acknowledgement}\\ 
We are indebted to Susanne Still and Matteo Marsili for collaboration and useful discussions years ago on joint works preceding the present one. Although they did not participate in this work, their ideas have remained a source of inspiration for us. I.K. is obliged to Risi Kondor for several enlightening discussions.

\bibliography{OptimizingESunderl1}
\end{document}